\documentstyle[psfig,epsfig,aps,multicol,fancyheadings]{revtex} % gallery format
\begin{document}
%\draft
%\psdraft
\title{Random manifolds in non-linear resistor networks: Applications
to varistors and superconductors}
\author{A. Donev, C.E. Musolff and P.M. Duxbury \footnote{ \noindent 
{\bf email:} duxbury@pa.msu.edu}, }

\address{ Dept. of Physics/Ast. and 
Center for Fundamental Materials Research,\\
Michigan State University, East Lansing, MI 48824, USA.}
\maketitle

\begin{abstract}
We show that current localization
 in polycrystalline varistors
 occurs on paths which are, usually, in the universality
class of the directed polymer in a random medium.  We also show
that in ceramic superconductors, voltage localizes
on a surface which maps to an Ising domain wall.  
The emergence of these manifolds 
is explained and their structure is illustrated using 
direct solution of non-linear resistor networks.

\end{abstract}


\begin{multicols}{2}[]
\narrowtext

Low energy paths and surfaces in random systems are of 
broad importance in statistical mechanics and in 
materials' theory.  Examples include the morphology of flux
lines in the presence of random pinning\cite{blatter94}, magnetic domain
walls in random-bond Ising magnets\cite{huse85,fisher86}, 
dielectric breakdown paths\cite{duxbury87,beale88}
and fracture surfaces\cite{bouchaud97}.  The directed polymer in a random medium
has received an enormous amount of interest, both due to 
its intrinsic importance and due to its relation with
growth processes\cite{kardar86}.  Domain walls in random-bond Ising models
have also received a good deal of attention, in fact the 
directed polymer in a random medium was first used to model
domain walls in two-dimensional random-bond magnets\cite{huse85}.
  More recently it has been realised that random 
path and surface problems map to  classic problems
 in computer science\cite{cormen90,ahuja93}.  In particular, the 
{\it shortest-path problem} is related to the directed
polymer in a random media\cite{marsili98} 
and the {\it minimum-cut problem} is
related to domain walls in random Ising magnets\cite{middleton95}.
The shortest path and minimum cut 
problems have polynomial time algorithms, and this has enabled 
physicists to solve path problems {\it with overhangs}
(i.e. not just the directed case)\cite{marsili98},
and domain wall problems in three dimensions\cite{middleton95}, both of which were
previously considered difficult.  Many more connections between
statistical physics and solvable combinatorial optimization problems 
have recently proven profitable\cite{alava01}.  Here we show that the shortest path
and minimum-cut problems also emerge in the study of the onset of current
or voltage in non-linear resistor networks and in their
applications to varistors and superconductors.  This enables a 
detailed characterization of the manifolds on which current
or voltage localize near critical thresholds.

Non-linear flow in random networks is germane to 
problems ranging from non-Newtonian fluid flow in porous
media\cite{sahimi98}, to the onset of flow in 
varistors\cite{bartkowiak96,vojta97,clarke99,heaney00}, and to the
onset of voltage in grain-boundary 
limited superconductors\cite{rhyner89,riedinger90,haslinger00}.
The emergence of special flow paths in the fluid and varistor
cases is well-established and network models
with realistic non-linear flow laws have 
been constructed\cite{sahimi98,bartkowiak96,vojta97}. 
Recently lattice models have also 
 been used to model the behavior of 
polycrystalline superconductors\cite{haslinger00}.
 The networks which model 
these materials have highly non-linear current-voltage
(or pressure-flow) behavior on each bond of a graph.
In the case of diodes or varistors each bond has a 
critical voltage, $v_c^k$, which varies from bond to bond
\cite{stinchcombe86,roux87,bartkowiak96,vojta97}. 
The onset of current flow occurs 
on special paths and the identification and
characterisation of these paths is 
a central issue in the analysis of varistors.
In the superconductor case, each
bond has a local critical current, $i_c^k$, below
which the bond carries current {\it but has
zero voltage}.  Above this threshold current, 
a local voltage must be applied to the
bond in order to increase the current further.
The onset of voltage occurs
 on a surface\cite{haslinger00} which
we show maps to a minimum cut through the
network.  

Firstly we define the shortest path and
minimum-cut problems\cite{cormen90,ahuja93}.
These problems (and the algorithms which solve them)
are very basic to algorithmic systems in computer science. 
 They are defined
on a graph where each edge is assigned a cost
(shortest-path problem) or a capacity (minimum-cut
problem).  Consider then a graph composed of 
nodes and edges, where each edge has a cost, 
$c_k$.  The shortest path between any two 
sites, $s$ and $t$, in the graph is simply the path on which
the sum of the edge costs, 
\begin{equation}
C_{st} = {\rm min}_{P}(\sum_{k\epsilon P} c_k),
\end{equation}
on the path is {\it smallest}.
In physics language the shortest path is the lowest 
energy path. It is clear that this is the same
as the problem of a directed polymer in a
random media, provided the path is forced
to be directed.   However if overhangs are allowed
this problem can also be solved efficiently using,
for example, Dijkstra's method. It turns out that
overhangs do not change the universality class of the
problem unless disorder is strong\cite{cieplak96,porto99}.  

The minimum-cut problem\cite{ahuja93,cormen90} is defined on
a graph consisting of nodes and edges
in which each edge has a flow {\it capacity},
$u_k$.  This is the maximum amount of flow that
the edge can carry.  There is no energy cost in
the computer science definition of this problem,
there is just the limiting capacity,
$u_k$ (However in the end some of the physics applications associate
the capacity with an energy).  Now consider the amount of flow that
can be carried between two sites, $s$ 
and $t$.  A key theorem states that
the maximum flow between $s$ and $t$ is equivalent
to the capacity of the minimum cut between $s$ and $t$\cite{ahuja93,cormen90}.
The minimum cut is the lowest capacity surface of separation in which
$s$ is on one side of the cut and $t$ is on the other. 
The capacity of the minimum cut, $U_{st}$, is the sum of the capacities on this
surface of separation, $S$, 
\begin{equation}
U_{st} = {\rm min}_{S}(\sum_{k\epsilon S} u_k).
\end{equation}
If we associate the bond capacities with
the exchange constants in a random-bond Ising magnet, then
the minimum cut maps to the lowest energy domain wall.
Somewhat surprisingly, the minimum cut can
be found efficiently (in almost linear time in the
number of edges) and this has enabled detailed 
study of domain walls in three dimensions\cite{middleton95,alava96}.
 
Now we consider in more detail the emergence
of the shortest path and the minimum cut in
non-linear random resistor networks.
To be concrete, we consider square networks
where each bond has a {\it monotonic} (single-valued)
 non-linear V-I  behavior.  
The case of current onset at a 
critical voltage is applicable
to varistors\cite{clarke99}(Fig. 1) and diodes  
and also to Bingham plastics\cite{sahimi98}
(replace voltage by pressure and current flow by fluid flow so
that flow initiates at a critical pressure).  The case of a voltage
onset at a critical current is applicable to 
ideal josephson junctions (Fig. 2) and
also to ideal flux-flow laws \cite{haslinger00}.  
Ideal varistor behavior is mathematically described by, 
$i(v) = (i_0 + {v-v_c\over r})\theta(v-v_c)$,
while ideal Josephson junction or flux flow behavior
is described by, $v(i) = (v_0 + {i-i_c\over g}) \theta(i-i_c)$,
where $\theta(x)$ is the step function which is 
zero for $x<0$ and one for $x\ge 0$.
 Finding the current or fluid flow through these 
non-linear networks seems  difficult 
 due to the non-smooth behavior at the critical
current or voltage.  However a variety of regularization
methods are available to handle these difficulties.
The simplest is to replace the step function by,
$\theta(x) = x/(\xi^2 + x^2)^{1/2}$ and to 
take the limit $\xi \rightarrow 0$.  This form
has nice analytic properties which assist in the
numerical analysis, so we have used this regularization
in our codes.
A key observation is that these problems have
a related cost function defined by,
$cost = \int_0^i v(i') di'$, which is 
{\it convex} provided the V-I law on each bond is
monotonically increasing.  With suitable
regularization this is true and solving these
non-linear resistor networks then  reduces to a convex
optimization problem with the constraint of flow
conservation at each node.  We have used this
framework to develop efficient codes for this
class of problem.  Here we discuss applications
to varistors and superconductors.

In the varistor case (see Fig. 1) each bond has
a random onset voltage, $v_c^k$, and the whole network has
the onset voltage, $V_c$.  At $V_c$,  current only
flows on the shortest path
through the network, as we confirmed numerically by finding the
path on which $V_c = \sum_{k\epsilon P} v_c^k$ is minimal (using Dijkstra's
method).   Voltage localization in the superconductor case is 
illustrated in Fig. 2.  In this case each bond has a random
critical current, $i_c^k$, and the whole network has the critical current,
$I_c$. We solved the
full flow equations directly and then confirmed that at $I_c$
voltage localizes on the {\it minimum cut}, i.e. the surface on which
 $I_c = \sum_{k\epsilon S} i_c^k $ is minimal.
   
The easiest way to prove that 
 random manifolds emerge 
at the macroscopic thresholds, $I_c$ and $V_c$,
 is by using the cost function, $\int_0^i v(i') di'$.
From this cost function,  it is evident that if either the current
or voltage can be kept to zero,
then the cost itself is zero.  In the case of a varistor, as the
external voltage is increased, the network 
distributes the voltage drops so as 
 to keep the current at zero.  It succeeds in doing this until
it is no longer possible and the first time this occurs
is on the shortest path, occuring at 
the critical voltage, $V_c$.  This argument applies to {\it any cost
function which has strictly zero current up to threshold}.  This
means that the behavior of the local $v-i$ characteristic after threshold
has {\it no influence} on the value of $V_c$.
In a similar manner, in the superconductor case, a large network
attempts to keep the voltage at zero in all of the bonds 
until it is impossible to do so.  As the applied current is
increased, the current is distributed so as to keep the currents
in all of the bonds below their threshold value, again in a very
co-operative manner.  The first applied current, $I_c$, 
at which voltage appears is determined by when a surface of bonds all
have reached their local critical currents.  This surface is the
minimum cut. 

To make contact with experiment, 
note that varistors are materials which are insulating below a critical
electric field, $E_c$, after which
 they become highly conducting\cite{bartkowiak96,vojta97,clarke99}.
  The physics of their
operation is understood to originate in the grain boundaries
of polycrystalline ceramics.  The interior of the grains of
typical varistors are conducting, but the grain boundaries
are insulating at low voltage (below about  3V). 
The onset of current flow occurs when 
the applied voltage is large enough to induce 
a path of grain boundaries to become conducting.  This physics
is naturally encapsulated in lattice models in which
a node represents a grain and a bond represents a grain boundary
\cite{bartkowiak96,vojta97,clarke99}.
However the grains have varying sizes and the grain boundaries
have varying properties, with some boundaries having higher
onset voltages than others.  That is, we must study
disordered networks.  These networks are also highly non-linear
as has been demonstrated by studies of flow onset 
across individual grain boundaries\cite{clarke99}.  
The linear dimension, $L$ of the lattice models is related
to the size of real varistors by $L = l/g$, where $l$ is the sample
size and $g$ the grain size.
Considerable progress has been made using numerical simulations
which have provided good agreement 
with experiment\cite{bartkowiak96,vojta97,clarke99}.  Here we make
the  connection with the statistical physics of random paths.

As discussed above, network models of polycrystalline varistors have
a current onset given by the minimum over all paths of 
$V_c = \sum_{k\epsilon P} v_c^k$.  This is just the
shortest-path problem of Eq. (1), where the voltage
threshold on a bond replaces the cost of that bond.
The shortest-path problem 
is usually in the universality class of the directed polymer in 
a random medium(DPRM)\cite{marsili98} 
(execept for strong disorder\cite{cieplak96,porto99}) 
so that $V_c$ behaves like the {\it energy} of a DPRM, i.e.
\begin{equation}
V_c = a_1  L  + a_2 L^{\theta},
\end{equation}
where $a_1$ and $a_2$ are independent
of sample size and the exponent
$\theta$ is universal.  Its value is known to 
be exactly $1/3$\cite{huse85} for paths through two dimensional
systems and $0.248 \pm 0.004$ for paths through
three dimensional media\cite{kim91}.  A key feature
of (3) is that the threshold electric field
of the lattice $E_c = V_c/L$ is {\it size
independent}, in contrast to the size effects
produced by rare-fluctuation theories of electrical
and dielectric failure\cite{duxbury87,duxbury94,beale88}, where $E_c$
approaches zero logarithmically in the large lattice limit.
   However the key
difference is that in dielectric breakdown, 
local regions irreversibly make the transition to the conducting
state, so that no voltage is required to maintain these
regions in the conducting state after failure.  In contrast,
 high quality 
varistors are reversible so that there is a
steady state current at a fixed applied voltage.  
The path on which current flows is usually self-affine and has length, $L_p$,
given by\cite{huse85,kim91},
$L_p = b_1 L + b_2 L^{\zeta}$,
where $b_1$ and $b_2$ are dependent
on the disorder distribution, but the exponent
$\zeta$ is universal.  Its value is known to 
be exactly $2/3$\cite{huse85} for paths through two dimensional
systems and $0.62 \pm 0.01$\cite{kim91} for paths through
three dimensional media.   The exponents
$\theta$ and $\zeta$ are related by,
$\theta = 2\zeta -1$, which is consistent with the 
numerical results quoted above.  
The difference in voltage between the lowest 
threshold path and the onset voltage of the next filamentary
current path scales in the same way as the energy
gap in the DPRM problem, which decreases logarithmically
with increasing sample size\cite{seppala01}.  In the
strong disorder limit, the paths become highly tortuous
and are no longer in  the DPRM class\cite{cieplak96,porto99,dobrin01}.

In the case of 
ceramic superconductors the low-angle grain
boundaries have much higher critical currents than
the high-angle grain boundaries\cite{dimos88}.  As in the 
varistor case, lattice 
models  take the grain centers to be nodes and the
links between grains to represent 
the grain boundaries\cite{rhyner89,haslinger00}.
Grain boundaries may  have a flux flow character
or a Josephson Junction character or be resistive. 
 If a grain boundary is resistive,
its critical current is zero.  As demonstrated above,
the onset of voltage occurs on the
minimum cut through such a network and this minimum cut
is related to domain walls in Ising magnets
\cite{huse85,fisher86,middleton95,alava96}.  The critical
current of these networks then  behaves in the same way as the
energy of Ising domain walls, i.e.,
\begin{equation}
I_c = c_1 L^{d-1} + c_2 L^{\theta}
\end{equation}
where $d$ is the spatial dimension, 
$c_1$ and $c_2$ are independent of sample size, and
$\theta = 1/3$ in two dimensions and $\theta = 0.82\pm 0.02$
in three dimensions.
There is thus a sample size independent critical current
density in these networks, in contrast to the size effect
which occurs in fuse networks\cite{duxbury87,duxbury94}. 
 Again this difference
is due to the fact that these networks must maintain a
current of at least $i_c^k$ on the $k^{th}$ bond
in order for a voltage to appear there.   The
surface on which voltage localizes at $I_c$ (i.e. the minimum cut)
 is in the universality class of domain walls in the 
random-bond Ising model and so it is self-affine
and has asymptotic roughness, $w$, given by, $w = c_3 L^{\zeta}$,
where $\zeta = 2/3$ in two dimensions and $\zeta \sim 0.41 \pm 0.01$
in three dimensions\cite{fisher86,middleton95,alava96}. 
The exponents $\theta$ and $\zeta$ are related to each other
by the scaling relation, $\theta = 2\zeta + d -3$.

We have demonstrated that current or voltage 
localize on special manifolds in non-linear networks.
As particular examples,  we showed that flow in a varistor 
begins on a path which is equivalent to a shortest
path through the  random medium, and in
 superconductors, voltage onset  occurs on a 
 minimum cut through the network.
The macroscopic current or voltage thresholds are 
then equivalent to the {\it  energy} of a random 
manifold.  Since these macroscopic thresholds only 
depend on the {\it  local current or voltage thresholds} 
(and not on the local  $v-i$ characteristics after threshold)
the manifolds on which current or voltage
localize in varistors and superconductors 
are described by the universal directed polymer
or random surface exponents.

This work has been supported by
the DOE under contract DE-FG02-90ER45418,
and by a subcontract through Sandia
National Laboratories.

%%%%%%%%%%%%%%%%%%%%%%%%%%%%%%%%%%%%%%%%%%%%%%%%%%%%%%%%%%%%%%%%%%%%%%%%%%%%%%%%
%% REFERENCES
%%%%%%%%%%%%%%%%%%%%%%%%%%%%%%%%%%%%%%%%%%%%%%%%%%%%%%%%%%%%%%%%%%%%%%%%%%%%%%%%

%***********************************************
\begin{figure}
\centerline{\epsfig{file=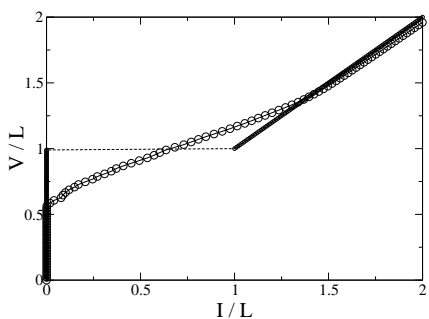,width=5cm,angle=270}}
\centerline{\epsfig{file=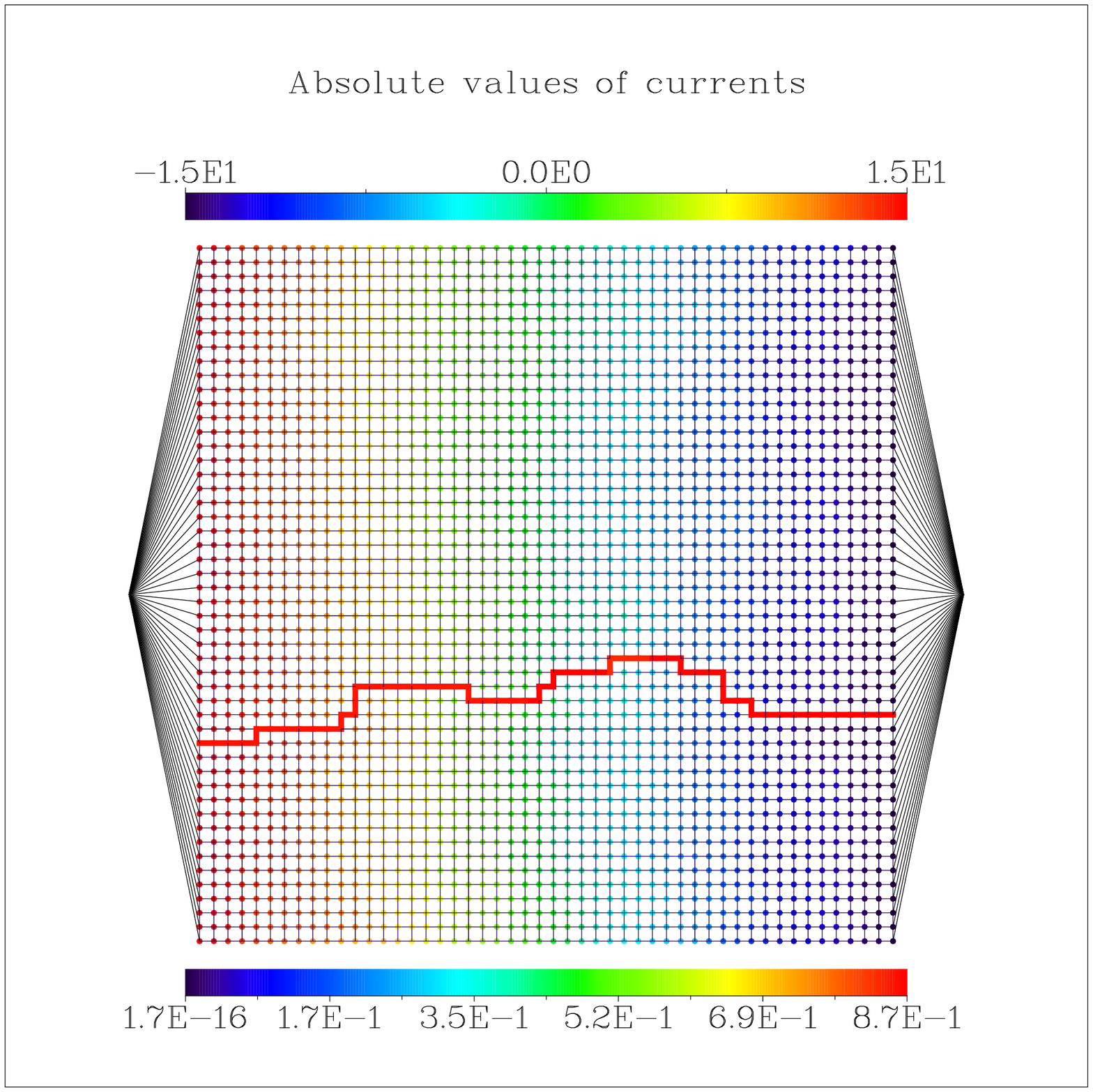,width=5cm,angle=0,
clip=TRUE,viewport=120 100 450 430}}
\vspace{0.1in}
\caption{Behavior of a  50x50 varistor network, with the
bonds having onset voltages uniformly distributed on the
interval [0,2],  asymptotic resistance one ohm, and with
the external voltage applied in the horizontal direction.  The
behavior of the average bond is the abrupt curve(smaller circles) in the 
top figure, while the behavior of the network is the smoother
curve in the top figure(bigger circles).  The localization
of current on the shortest path is illustrated in the lower
figure.  This is the current pattern at $V/L = 0.59$.  At higher
voltages further filamentary current paths emerge.}
\label{fig1}
\end{figure}

%***************************************************
\begin{figure}
\centerline{\epsfig{file=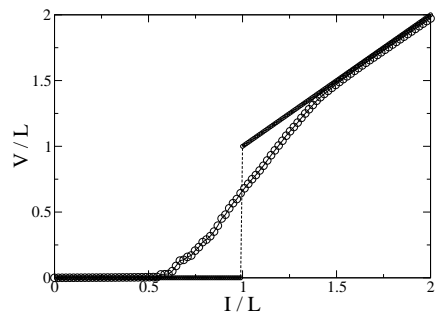,width=5cm,angle=270}}
\centerline{\epsfig{file=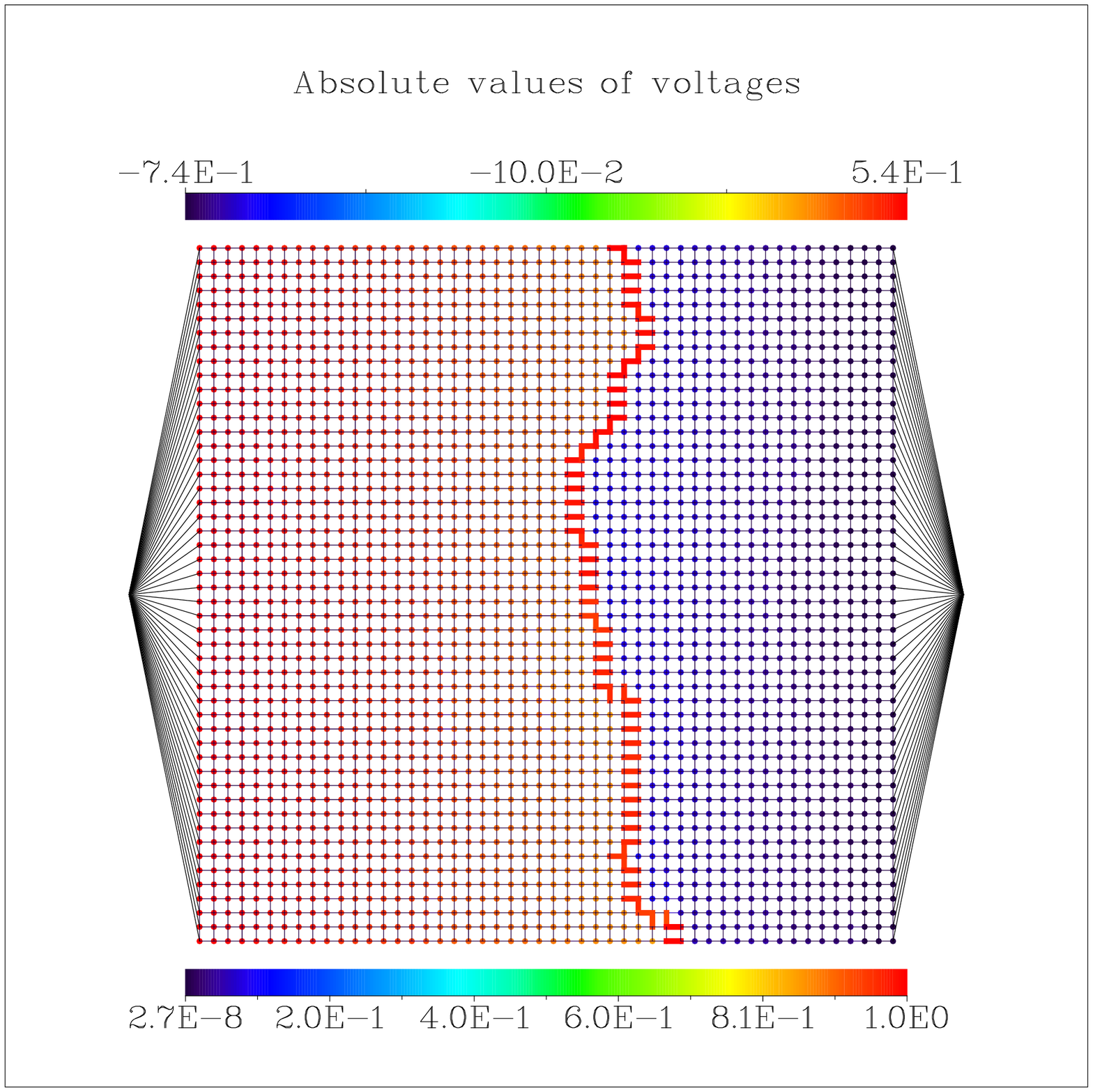,width=5cm,angle=0,
clip=TRUE,viewport=120 100 450 430}}
\vspace{0.1in}
\caption{Behavior of a 50x50 Josephson junction network, with the
bonds having onset currents uniformly distributed on the
interval [0,2] and asymptotic resistance one ohm, and with 
the external current injected in the horizontal direction.  The
behavior of the average bond is the abrupt curve(smaller circles) in the 
top figure, while the behavior of the network is the smoother
curve in the top figure(bigger circles).  The localization
of voltage on the minimum cut is illustrated in the lower
figure.  This is the voltage pattern at $I/L = 0.58$.  At higher
currents further sheet-like voltage surfaces emerge. }
\label{fig1}
\end{figure}

\end{multicols}

\end{document}